\documentclass[pdflatex,sn-mathphys-num]{sn-jnl}% Math and Physical Sciences Numbered Reference Style

\usepackage{graphicx}%
\usepackage{multirow}%
\usepackage{amsmath,amssymb,amsfonts}%
\usepackage{amsthm}%
\usepackage{mathrsfs}%
\usepackage[title]{appendix}%
\usepackage{xcolor}%
\usepackage{textcomp}%
\usepackage{manyfoot}%
\usepackage{booktabs}%
\usepackage{listings}%

\newcommand{\ket}[1]{\left| #1 \right>} % for Dirac bras
\begin{document}

\title{Bell's theorem: why probability factorisation fails}

\author*{\fnm{Geoff} \sur{Beck}}\email{geoffrey.beck@wits.ac.za}

\affil{\orgdiv{School of physics and centre for astrophysics}, \orgname{University of the Witwatersrand}, \orgaddress{\street{1 Jan Smuts Avenue}, \city{Johannesburg}, \postcode{WITS-2050}, \state{Gauteng}, \country{South Africa}}}

\abstract{The empirical proof of Bell inequality violations was a landmark moment for research into quantum foundations. It commits us to a universe without strict relativistic locality or requires that we escape through a potential loophole like many-worlds or superdeterminism. In this work, we consider a sequential, single-spin experiment whose auto-correlations match the two-spin entangled correlations of a Bell scenario. We use a counterfactual equivalence between the two to argue that Bell-type correlations are actually just a different implementation of the sequential experiment. This comports well with the counterfactual basis of the original EPR argument, and explains any apparent non-locality as a consequence of indirectly measuring quantities that do not have predefined values, due to the state altering nature of sequential spin measurements. }

\keywords{Bell's theorem, Bell inequalities, quantum foundations, non-locality}

\maketitle

\section{Introduction}
The outcome of the the Einstein-Podolsky-Rosen~\cite{epr1935} experiment is perhaps one of the most shocking results to emerge from quantum mechanics. Actual tests were performed via the EPR-Bohm version~\cite{bell_einstein_1964}, but still yield apparent evidence~\cite{aspect1982} that reality is non-local~\cite{maudlin2014,sep-bell-theorem}. This non-locality takes a form whereby the outcome of spacelike-separated experiments are interdependent, yet does not allow for actual signalling (as the outcomes cannot be controlled). Thus, our non-locality is conveniently set up to preserve relativity in practical terms (no signalling), even while flagrantly violating its spirit. In spite of this, ``no signalling'' remains insufficient to derive the observed quantum correlations~\cite{1994FoPh...24..379P}. 

Bell's theorem~\cite{bell_1964} considers two systems at spacelike separation and constructs an inequality out of their correlations that should, in principle, be obeyed by classical theories. This is because the inequality assumes that the distant systems obey ``local causation'', we are free to choose the experimental parameters, and each measurement generates a single outcome. In both practice and theory, quantum mechanics has been found to violate this inequality when such spacelike separated systems are entangled. This means we must either relinquish local causation or appeal to something like superdeterminism~\cite{superdeterminism}, or many-worlds~\cite{many-worlds}. In this work we will not be concerned with these loopholes. Our focus is thus confined to the local causality condition, which can be considered as the conjunction of three elements~\cite{Cavalcanti_2014,sep-bell-theorem}
\begin{enumerate}
	\item Relativistic Causal Structure (RCS)
	\item Probability Completeness (PC)
	\item Reichenbach's Common Cause Principle (RCCP)
\end{enumerate}
The first point can be thought of as signifying that both events occur in the same Lorentzian Spacetime (LS) and that causation is consistent with this spacetime or Consistent Lorentzian Causation (CLC). The latter means that causes precede effects, and causal relationships are not possible at spacelike separation. 
Condition 2, or PC, refers to the idea that a set of mutually exclusive events ${x}$, where the sum of probabilities $\sum P(x) = 1$, satisfies
\begin{equation}
	P(y) = \sum_x P(x) P(y|x) \; ,
\end{equation}
for any $y$. Here $P(y|x)$ is the conditional probability of $y$ given that $x$ holds, this is given by
\begin{equation}
	P(y|x) = \frac{P(x,y)}{P(x)} \; ,
\end{equation}  
where $P(x,y)$ is the probability of both $x$ and $y$ occurring or $P(x\cap y)$.
Condition 3, Reichenbach's common cause principle, can be expressed as requiring two elements: Common-Cause Screening (CCS) and Probability Factorisation (PF)~\cite{Cavalcanti_2014}. The first is that if two events $A$, $B$ are correlated, or $P(A,B) > P(A)P(B)$, then there are only three possible, mutually exclusive, causal relations: $A$ causes $B$, $B$ causes $A$, or some $X$ causes both $A$ and $B$. The condition of Probability Factorisation (PF) is given by
\begin{equation}
	P(A,B|X) = P(A|X,B) P(B|X) =  P(A|X) P(B|X) \; . \label{eq:non-sep}
\end{equation}
Here we can envisage $X$ as the ``common cause'' of both events $A$ and $B$. The RCCP uses the tool PF to determine where CCS is present and allows us to identify common causes of correlated events. 

Thus, we can imagine the following logical relationship: RCS $\cap$ PC $\cap$ PF $\cap$ CCS $\implies$ local causality. We can name this relationship as Bell's Condition of Local Causation (BCLC). All constituent conditions are necessary to Bell's condition of local causation, i.e. local causation is believed to fail if any of them fail. In contrast, none of the conditions are independently sufficient to justify local causation. 
% We can express this as 
% \begin{align}
% 	\neg c_i  & \implies \neg\mathrm{BCLC} \; , \\
% 	\neg (c_i & \implies \mathrm{BCLC}) \; , 
% \end{align}
% where $\neg c_i$ indicates condition $i$ is false, where $c_i \in$ \{RCS,PC,PF,CCS\}. 
A violation of Bell's inequality will thus necessitate renouncing at least one of these four elements. 

We can renounce LS by augmenting the Lorentzian structure of spacetime, e.g. \cite{Durr_1999,maudlin1994}. One can also opt to violate CLC by introducing retro-causality, see \cite{sep-qm-retrocausality} for a review. The conditions of probability completeness (PC) or factorisation (PF) seem quite abstract in contrast, as they do not directly suggest new or modified physics. In this work we will be interested in the interplay of PF and Common-Cause Screening (CCS), but see \cite{hofer2012,hofer2013} for details on an approach that violates PC instead. 

Dropping or modifying the RCCP has been quite commonly advocated in the literature, as many authors~\cite{vanFraassen1982-VANTCO-63,Fine1989-FINDCN,Elby1992-ELBSWE,butterfield1992,leifer_spekkens_2013} find the other options unpalatable. Notably, dropping LS can feel rather like reintroducing a Lorentz aether reference frame, in that one is adding some kind of privileged and unobservable structure to explain puzzling observations while evading the obvious successes of relativity. Retro-causation faces rather severe obstacles due to questions over its coherence, see e.g. \cite{Mellor1998-MELRTI} or fine-tuning~\cite{Wood_2015}. The idea that PC should be dropped seems very unintuitive, and does not even achieve the desired results in the case presented by \cite{hofer2012,hofer2013}. According to the analysis of \cite{Cavalcanti_2014}, a Bell scenario results in any product-state from the entangled systems' common past satisfying the common cause requirement from \cite{hofer2012,hofer2013}. This leaves us with renouncing the conditions PF or CCS. If we drop CCS this introduces some physical influence that also violates either LS or CLC, since the two correlated, spacelike-separated events depend on each other as well as their common past. %Notably, \cite{butterfield1992} proposes that if we exclude superluminal causes we must renounce CCS, although the author uses a weakened form of this condition. 

There still remains the possibility of dropping probability factorisation instead, this is done in the literature by \cite{leifer_spekkens_2013} as part of a project aimed at interpreting quantum mechanics as a scheme of Bayesian inference. Proposals which violate or replace the RCCP, e.g. \cite{hofer-szabo199,Unruh2002-UNRIQM,Redei2002-RDERCC-4,leifer2006,hofer2012,hofer2013,leifer_spekkens_2013}, face the difficulty of explaining correlations without it, as discussed in \cite{Cavalcanti_2014}. This is because one no longer has a means of identifying common causes. In addition, using the quantum state as common cause for Bell correlations faces the difficulty that these depend upon measurement settings that may be selected after preparation of the state~\cite{Cavalcanti_2014}. %Notably, some authors question whether Bell correlations can/should be explained causally at all~\cite{Fine1989-FINDCN,Elby1992-ELBSWE}. 

% It is stated in \cite{Brunner_2014} that PF represents a ``precise condition for locality in Bell experiments''. This seems sensible, as it amounts to requiring that the events $A$ and $B$, with common cause $C$, have no dependence on each other when they are space-like separated. Thus, PF seems to constitute a necessary condition for locality. To examine this more closely it is worth noting that Bell inequalities are realisable within systems of classical optics, see e.g. \cite{classical0,classical1,classical2,classical3,classical4,classical5}. The unifying theme of these examples is one of the violating probability factorisation, so that Eq.~(\ref{eq:non-sep}) does not hold. These systems clearly involve no violation of actual local causation, only a condition analogous to BCLC. The conclusion is that PF is only a sufficient, but not necessary, condition for actual local causation. This indicates that a failure of local causation would only be a sufficient condition for Bell inequality violation, rather than necessary to it, as we can violate PF without affecting the status of local causation. One might argue that this is not so clear in the case of an actual quantum mechanical Bell experiment, where we deal with two entanglement systems in spacetime (rather than correlated properties of the same classical system that violate PF). Thus, we must reinforce our argument with explicit analysis of the causal links in a Bell-type scenario.  

It is worth noting that quantum mechanics itself violates PF. The RCCP then implies some causation between Bell measurements, but one that does not have a time order. Hence, we must violate locality anyway. Another approach to the PF problem would be to show that violating PF is not a faithful sign of causation. Indeed, Bell-type inequalities can be violated in a variety of classical systems via PF~\cite{classical0,classical1,classical2,classical3,classical4,classical5,classical6}. However, none of these involve a system that is inseparable across physical space. Our strategy is instead to focus on a similar kind of system, a single spin undergoing two sequential polarisation measurements. Remarkably, these experiments have nearly identical correlations to entangled Bell states (differing only in a sign). Further, we argue that there is a counterfactual equivalence between these two classes of experiment. This takes the form of using the entangled pair to reconstruct the sequential auto-correlations for a single spin measured twice on different axes. The role of entanglement is then merely to facilitate a counterfactual measurement. The failure of PF thus emerges from the reconstruction of a causal quantity, not some inherent property of entanglement. We further motivate this idea by demonstrating that the difference between a spin product state and an entangled one is the latter contains only terms that respect a spin matching condition. That is, the same correlations would arise between a pair of completely unrelated spins that happened to have highly correlated responses on all axes. This demonstrates that the counterfactual/matching relationship between spin pairs is alone sufficient to violate PF. 
%We then address the GHZ experiment~\cite{ghz_1990}, showing its spin correlations similarly admit a sequential equivalent. 

In effect, we disarm Bell's ``locality bomb'' by arguing that a Bell experiment is just a different implementation of sequentially measuring spin auto-correlation on differing axes. To complete this, we demonstrate that the surprising dependence of Bell violations on 3-spin correlations in Fine's theorem~\cite{fine_hidden_1982} is only explicable via our counterfactual equivalence. Thus, any shocking ``non-local'' results can be laid at the feet of a purely local property being measured surprisingly. This solution is attractive because it requires only a commitment to realism (via EPR's ``elements of reality''~\cite{epr1935}). However, it leaves open the problem of what local hidden variables are missing to be able to emulate quantum results without non-locality.

This paper is structured as follows. In section~\ref{sec:bell} we lay out the basic requirements of Bell's theorem. Then, in section~\ref{sec:counterfactual} we explore the linkage between Bell's entangled spin scenario and sequential measurements on single spins. Next, we consider the role of spin matching conditions in section~\ref{sec:pairs}.
%We broaden our examination to the GHZ scenario for inequality free Bell tests in section~\ref{sec:ghz}. 
In section~\ref{sec:fine} we demonstrate that this relationship explains an otherwise puzzling aspect of Fine's theorem, and then draw conclusions in section~\ref{sec:conclusion}.

\section{Bell correlations}\label{sec:bell}
In a Bell experiment we prepare a state of two entangled spins
\begin{equation}
	\ket{\psi} = \frac{1}{\sqrt{2}}\left(\ket{\uparrow\downarrow} - \ket{\downarrow\uparrow} \right) \; .
\end{equation}
These spins are subsequently measured at sites $A$ and $B$ which are spacelike separated. Crucially, $A$ and $B$ can measure their spins on different axes. The correlation between $A$ and $B$ measurements is given by
\begin{equation}
	C_{A B} (\theta,\phi) = \langle \sigma_{\theta,A} \sigma_{\phi,B}\rangle = - \cos(\theta-\phi) \; , \label{eq:bell}
\end{equation}
where $\theta$ and $\phi$ are physical orientations of the measurement devices. 
Bell's theorem then states that~\cite{bell_einstein_1964}
\begin{equation}
	-1 \leq C_{A B} (\theta_1,\theta_2) + C_{A B} (\theta_1,\theta_3) - C_{A B} (\theta_2,\theta_3) \leq 1 \; ,
\end{equation}
for theories that obey Bell's local causation condition. 
% Crucially, causal influences are judged following Reichenbach's principle of common cause, in the form that non-separable probability statements
% \begin{equation}
% 	P(A,B|X) = P(A|B,X) P(B|X) \neq P(A|X) P(B|X) \; ,
% \end{equation}
% imply a causal relationship between events $A$ and $B$ that is not sufficiently explained by a common cause in $X$.

In particular, the spectre of non-locality arises from quantum probabilities for the outcomes at $B$, or $B_\uparrow$ and $B_\downarrow$, retaining a dependence on those of the spin at $A$
\begin{equation}
	P(B_\uparrow|A_\downarrow,\psi,\theta,\phi) = \cos^2\left(\frac{\phi}{2}-\frac{\theta}{2}\right) \; , 
\end{equation} 
which is not equal to $P(B_\uparrow|\phi,\psi) = \frac{1}{2}$.

\section{Sequential spin measurements and counterfactuals} \label{sec:counterfactual}
The dependence of outcome probabilities on the other spin deserves some deeper examination. Particularly as our Bell state contains strong anti-correlations between $A$ and $B$ outcomes. For instance, if $A$ and $B$ use the same axis of measurement, they always get opposite results. For the sake of clarity, we will explicitly label our outcomes by the axis of measurement, e.g. $P(B_{\uparrow,\phi}|A_{\downarrow,\theta},\psi,\theta,\phi)$, and note that
\begin{equation}
	P(B_{\uparrow,\phi}|A_{\downarrow,\theta},\psi) = P(B_{\uparrow,\phi}|B_{\uparrow,\theta},\psi) \; , \label{eq:equiv}
\end{equation}
by virtue of the anti-correlations inherent in $\psi$. Here, of course, we are assuming the prepared spin values in the Bell scenario are ``elements of reality''. If one were to ask for a quantum observable that could access $P(B_{\uparrow,\phi}|B_{\uparrow,\theta},\psi)$, it would be natural to measure a spin $B$ on the $\theta$ and then $\phi$ axes sequentially, i.e. $\langle\hat{\sigma}_{\phi,B}\hat{\sigma}_{\theta,B}\rangle$. This yields a correlation function 
\begin{equation}
	\begin{aligned}
	C_{BB}(\theta,\phi) = \ &  P(B_{\uparrow,\phi}|B_{\uparrow,\theta},\psi) + P(B_{\downarrow,\phi}|B_{\downarrow,\theta},\psi) \\& - P(B_{\uparrow,\phi}|B_{\downarrow,\theta},\psi) - P(B_{\downarrow,\phi}|B_{\uparrow,\theta},\psi) \; ,
	\end{aligned}
\end{equation} 
which is equal to $-C_{AB}(\theta,\phi)$.

Remarkably, for a large class of spin density matrices we find
\begin{equation}
	C_{BB}(\theta,\phi) = \mathrm{Tr}(\hat{\sigma}_\phi\hat{\sigma}_\theta\hat{\rho}) = \cos(\phi - \theta) \; . \label{eq:seq}
\end{equation} 
Once again, the angles in question are the physical filter orientations, so that $\hat{\sigma}_\theta = \hat{R}\left(\frac{\theta}{2}\right) \hat{\sigma}_z \hat{R}^\dagger\left(\frac{\theta}{2}\right)$ with $\hat{R}$ being a rotation operator, and $\hat{\sigma}_z$ being the conventional Pauli operator. Note that this correlation holds for density matrices with no polarisation in the $\vec{y}$ direction. For states that do not adhere to this, an additional term $\propto i \sin(\phi - \theta)$ is included. This imaginary addition seems to reflect the polarisation in the perpendicular direction but does not modify the real correlation in the $x$--$z$ plane. Interestingly, the correlation in Eq.~(\ref{eq:seq}) is order independent, even though the operators do not commute. Note that the equivalence in Eq.~(\ref{eq:equiv}) even explains the sign difference between Bell and sequential correlations. 

Not only is our sequential measurement naturally linked to Bell scenarios via probabilities in Eq.~(\ref{eq:equiv}), its correlation function matches, allowing it to violate a Bell/CHSH inequality. Bell himself noted such measurements could produce a violation~\cite{bell_speakable_2004}, but considered it irrelevant, as he felt it can be easily explained by local hidden variables. 

Why can our sequential experiment violate an inequality all local scenarios should obey? The answer is the non-separable probabilities like $P(B_{\uparrow,\phi}|B_{\uparrow,\theta},\rho,\theta,\phi)$. These alone are sufficient to generate a Bell violation~\cite{sep-bell-theorem}. This non-separability does not seem obviously pathological to local causation. The order independence is an oddity, but stems only from the properties of the initial state (as the operators themselves do not commute in general). Moreover, any contextuality involved is trivial, in the sense that filtering measurements physically alter the state. Could this also be true for the entangled case? 

First, let us reconsider the Bell scenario and the role of the two measurements. Bell state anti-correlations make it clear that the operation at $A$ can act as a counterfactual measurement for spin $B$, i.e. it indicates the result we would have got, had we measured $B$ along the same axis as $A$ (if we maintain a commitment to realism). This is the role entanglement plays, it allows us to perform this counterfactual measurement. By measuring $A$ and $B$ on different axes, we can reconstruct the auto-correlations of $A$ or $B$ between two axes, in keeping with the original EPR argument~\cite{epr1935} for position and momentum. The special ingredient of non-separability/entanglement in a Bell state guarantees anti-correlation on all axes. Without this, a counterfactual measurement on an arbitrary axis is impossible. 

The apparent causal effect between measurements can now make sense. We are measuring the auto-correlation of a single spin, which does indeed depend causally on the angular difference between the measured axes. This naturally translates into the entangled-pair equivalent, without requiring the spacelike-separated measurements be causally related. That is, to the extent that they appear causal it is only because they are used to reconstruct a response that is causal. All of this makes it clear that a Bell experiment is not about the ``beables'' of the system, we are instead using the beables of the entangled pair to reconstruct an emergent result of sequential measurement determined by both initial state and apparatus settings. 

The equivalence between sequential and entangled experiments has been noted before and is not limited to spin correlations either, as demonstrated by Klyshko's advanced wave formalism in optics~\cite{klyshko_simple_1988}. An optical realisation of a Bell experiment in said formalism would be precisely the two sequential polarisers. Additionally, the similarity of Eqs.~(\ref{eq:seq}) and (\ref{eq:bell}) is noted in \cite{hess_2025}, where the authors state that ``mathematically there exists almost no difference between this Malus-type experiment and the EPRB-type as soon as Einstein's hypothesis of elements of physical reality is made.'' Indeed, this is embodied in Eq.~(\ref{eq:equiv}), as this requires something like elements of reality to validate the counterfactual. The novel addition here is the explicit demonstration of the consistency of the counter-factual. We further claim physical equivalence on this basis. Support for this can found in the original EPR thought experiment. Here, the position of one particle and the other's momentum can be used to reconstruct both quantities for both particles, despite the non-commuting nature of the observables (the uncertainty principle is still obeyed at the statistical level, as required). This is precisely what we have argued is happening in EPR-Bohm.    

% It is well known that spin measurements change the state being measured. How then could two sequential operations match a case of two separate ones, even if the subjects exactly match? The solution is hinted at by the order invariance of the sequential correlation. Filter order does indeed matter if we extend to further sequential measurements which depend on terms like $\cos(\theta_1 - \theta_2 + \theta_3)$. This occurs because the effects of each filter are not disjoint (they do not erase all previous correlations). However, this does not trouble us at only 2 filters, because the first filter instils correlations on the input state which are revealed by the second.  Our measurements here faithfully separate components on the filtered axis, only other directions are altered. So effects of compounding correlations would only be revealed by a third filter. In the entangled case then, we reveal the filter effects by correlating the results from matched spins on two different axes. The compounding disturbance of differing measurement orders played no role in the sequential case, so it is not missed here.

\section{A parable from EPR}\label{sec:pairs}
We can provide an alternative perspective by first constructing a fallacious demonstration of non-locality using the original EPR scenario. Here we have $A$ measuring a particle's position $\hat{q}_A$ while $B$ measures momentum $\hat{p}_B$. We gather statistics and compute standard deviations $\langle \Delta \hat{q}_A \rangle$ and $\langle \Delta \hat{p}_A \rangle$ (the second by inference from correlation). Shockingly, we find $\langle \Delta \hat{p}_A \rangle \cdot \langle \Delta \hat{q}_A \rangle \geq \frac{\hbar}{2}$. Surely, this implies measuring $\hat{p}_B$ has disturbed the state of particle $A$! 

Of course, we can debunk this by noting that if we consider a coherent state (for simplicity) and measure $\langle \Delta \hat{q} \rangle$ and $\langle \Delta \hat{p} \rangle$ in separate experiments then the uncertainty bound still holds by definition (see e.g. \cite{rae2002}). Thus, we have not demonstrated non-locality via our EPR measurements, despite appearances to the contrary. This is very strongly analogous to what we are arguing in this work, as the measure of connectedness is what failed us. Now, can we construct an even more precise analogy for pairs of spins?

Let us consider two devices that emit independent spins in the state $\propto \ket{\uparrow_z} + \ket{\downarrow_z}$. We then measure the spins on angles $\theta$ and $\phi$ from the $\vec{z}$ axis. We know that the probabilities for equal and opposite results are
\begin{align}
	P_\mathrm{eq} = \cos^2\left(\frac{\theta}{2}\right)\cos^2\left(\frac{\phi}{2}\right) + \sin^2\left(\frac{\theta}{2}\right)\sin^2\left(\frac{\phi}{2}\right) \; , \\
	P_\mathrm{diff} = \cos^2\left(\frac{\theta}{2}\right)\sin^2\left(\frac{\phi}{2}\right) + \sin^2\left(\frac{\theta}{2}\right)\cos^2\left(\frac{\phi}{2}\right) \; .
\end{align}
Thus, we can express
\begin{equation}
	C(\theta,\phi) = \cos(\phi-\theta) -4 \cos\left(\frac{\theta}{2}\right)\cos\left(\frac{\phi}{2}\right)\sin\left(\frac{\theta}{2}\right)\sin\left(\frac{\phi}{2}\right) \; . \label{eq:product}
\end{equation}
Interestingly, this does contain the Bell correlation, but suffers from a correction that prevents inequality violation. 

Let us examine the product state to determine why it cannot violate a Bell inequality. We have
\begin{equation}
	\ket{\psi} = \ket{\psi_1} \otimes \ket{\psi_2} \; ,
\end{equation}
or
\begin{equation}
	\begin{aligned}
		\ket{\psi} & = \frac{1}{\sqrt{2}}\left[\left(\cos\left(\frac{\phi-\theta}{2}\right) -\sin\left(\frac{\phi+\theta}{2}\right)\right)\right.\ket{\uparrow_\theta\uparrow_\phi} + \left(\cos\left(\frac{\phi+\theta}{2}\right) -\sin\left(\frac{\phi-\theta}{2}\right)\right)\ket{\uparrow_\theta\downarrow_\phi} \\ & + \left(\cos\left(\frac{\phi+\theta}{2}\right) +\sin\left(\frac{\phi-\theta}{2}\right)\right)\ket{\downarrow_\theta\uparrow_\phi} + \left.\left(\cos\left(\frac{\phi-\theta}{2}\right) + \sin\left(\frac{\phi+\theta}{2}\right)\right)\ket{\downarrow_\theta\downarrow_\phi} \right]\; .
	\end{aligned}
\end{equation}
Consider the coefficient $\cos\left(\frac{\phi-\theta}{2}\right) -\sin\left(\frac{\phi+\theta}{2}\right)$. The first term of this depends only on the angular difference, whereas the second is sensitive to actual values. This pattern is repeated throughout $\ket{\psi}$. If we selected only terms dependent on $\phi-\theta$ we are left with a Bell state. It is thus the presence of terms dependent on actual, rather than relative, values of $\theta$ and $\phi$ that prevent Bell violation (exactly as it appears in Eq.~(\ref{eq:product})). 

This becomes interesting if we consider where such a requirement could come from. A simple example is where the spins are linked by some conservation requirement, i.e. total angular momentum is fixed on all axes. However, the scenario itself doesn't matter, the key requirement is a definite relationship between outcomes if the measurement angles match, i.e. the spins either match or are opposite. This naturally requires that outcome probabilities must exclusively depend on $\phi -\theta$. In turn, such a dependence creates a Vorob'ev cyclicity~\cite{vorobev_consistent_1962} in the expectation values $C(\theta_i,\theta_j) \equiv C(\theta_i-\theta_j)$ when $i\neq j \ , i,j \in {1,2,..,n}$ with $n\geq 3$, which ensures the tripartite joint probability, e.g. $p(s_1,s_2,s_3)$ for spin outcomes $s_i = \pm 1$ on $\theta_i$, cannot be non-negative. Further, Fine's theorem~\cite{fine_hidden_1982} demonstrates the failure of the tripartite joint distribution to exist is both necessary and sufficient for a Bell violation in bipartite correlations.

Thus, we must conclude that any two sets of pairwise-matching spins must violate a Bell inequality. Regardless of how they are produced or whether they are entangled. Notably, we had to cull terms by hand from our product state above. This should be expected as we cannot set up an experiment where we ensure independent, non-interacting spins match on more than one set of axes, at least to the authors' knowledge. Indeed, perhaps entanglement is the only practical means of achieving such matching. However, since all that matters is the $\phi-\theta$ dependence, we can still argue that only the matching of the spin states is required in principle for Bell violation via PF.

Is the matching condition alone enough? We can construct a classical matching problem to check. Consider a set of tabulated data in two columns: $A$ and $B$. Entries can only take values $1$ and $0$. Whenever $A = 1$ we find $B = 0$ and vice versa. However, we also find that $P(A=1) = \frac{1}{2}$. Thus, we have $P(A=1|B=0) = 1$ and $P(A=1,B=0) = \frac{1}{2}$, whereas $P(A=1) P(B=0) = \frac{1}{4}$. In this case we have $P(A=1,B=0) > P(A=1) P(B=0)$ and also clearly do not have PF. Thus, $A$ and $B$ should be causally related. However, we can realise this with the following apparatus: a machine that deals out two coins which always show opposite faces. Clearly, there is a common cause, yet our condition to determine CCS has just failed. Exactly like the Bell scenario, the failure is tied to a matching condition. 

However, such a device cannot violate a Bell inequality. In this case we cannot define an analogue of mismatched measurement axes. If we chose a pair of matched classical dipoles we also would fail to violate the inequality, as a definite geometric relation exists between each and every projection. We cannot define a probability that both spins match, this is fully determined by the choice of angles (even if the initial state is uncertain). We cannot achieve a Vorob'ev cyclicity without probabilities. In the spin case, we have two possibilities on each axis, with only the bias determined geometrically. The peculiar ingredient a spin possesses is quantisation. Something quantised on all axes cannot use the orthogonal outcomes $s_x$, $s_y$, $s_z$ to determine the response on any arbitrary axis. Thus, in our classical dipole case, there is only one possible correlation between each axial pairing and this is fully determined by geometry. For the quantum case, there are four separate correlations, each of which is fully determined by geometry, but the realised outcome is not determined solely by the geometry.  

It is evident then why ``hidden variable'' models fail to reproduce quantum mechanics. Presupposition of a vector-like description of the spin, where outcomes on different axes share a single determined relationship, cannot account for quantisation. This is not ameliorated by a Bell-style model of $\mathrm{sign}(\vec{n}\cdot\vec{\lambda})$ for a spin with hidden state $\vec{\lambda}$ measured on an axis $\vec{n}$. As the sign of a given axis remains determined by the geometric relationship alone. The missing ingredient is not non-locality nor lack of predefined values, it is quantisation itself. Our argument can be reinforced by the fact that inter-axis correlations for a single spin show the same behaviour as the Bell state, with quantisation being the only possible culprit. 

We can also see an interpretation of what operator non-commutation means: the observable quantities cannot have non-statistical relationships. This provides a robust correspondence with a lack of common eigenstates. In this case, a spin prepared as $\ket{\uparrow_z}$ cannot return $\uparrow_\theta$ every time. A purely statistical correlation between the observables is demanded by their non-commutation. In this case, the physics come down to filtering nature of spin measurement/preparation acting on a quantised property. For material spins, the filtering magnetic field introduces a biased precession on misaligned axes.

\section{A Fine solution}\label{sec:fine}

So far, we have demonstrated that our interpretation is consistent. We can go further by also arguing it is the only way to consistently accommodate Fine's theorem~\cite{fine_hidden_1982}, which implies satisfying a Bell/CHSH inequality requires the existence of a 3/4-spin joint probability with marginals consistent with the 2-spin entangled case. For example, the original Bell inequalities can be derived by demanding the joint
\begin{equation}
	p(s_1,s_2,s_3) = \frac{1}{8} \left(1 + \sum_i B_i s_i + \sum_{i<j} C_{i j} s_i s_j + D s_1 s_2 s_3\right) \; , 
\end{equation}
is non-negative~\cite{halliwell_two_2014}. Note $s_i = \pm 1$ represents a spin measured on some choice of axis $\theta_i$, $B_i$ are the single-spin averages, $C_{i j}$ are the 2-spin correlations, and $D$ is that for 3 spins. It is quite surprising that an inequality built on 2-spin correlation measurements actually depends on a 3-spin correlation. Something that has no link to the experimental design and no obvious definition. To illustrate, $C_{i j}$ are constructed as the entangled 2-spin correlations. How then is $D$ defined? The only unambiguous definition would be for an entangled triplet. However, this doesn't seem to have any relevance to whether an entangled pair share non-local correlations.  

We cannot merely dismiss this curiosity, as Fine's theorem provides a condition that is both necessary and sufficient for an inequality violation, unlike Bell's own conditions. This means all violations come down to $p(s_1,s_2,s_3)$ failing to exist. Any actual physical lessons to be learned must therefore be drawn from this failure. It is thus of great interest to determine why the physically ambiguous $D$ is so vital to Bell violations. 

The relationship between single and entangled spin experiments in section~\ref{sec:counterfactual} provides us a reason the 3-spin correlation function appears in Fine's theorem. This function has no obvious or relevant definition in terms of entangled pairs. However, it is easily definable in the single spin consecutive experiment, we just add a third spin measurement. Moreover, such a sequential experiment is order dependent, meaning that $p(s_1,s_2,s_3)$ has no unique definition. We also note that the third filter is the first that reveals the effects of compounding disturbances. To illustrate this, we can consider Malus' law for 3 filters. Here we obtain different intensities than those predicted for quantum mechanics, as our quantum $D$ term depends only on functions like $\cos(\theta_1 - \theta_2 + \theta_3)$, which is not compatible with Malus' law and thus requires compounding filter effects. 

As an addendum, note that the 2-spin terms $C_{i j}$ are Malus-compatible. Establishing this requires an experiment using polarising beam splitters. An unpolarised beam enters such a PBS (oriented at angle $\theta_1$) and both output beams are subject to a second PBS aligned along $\theta_2$. We then label the divided beam components as $I_{+}$ (transmitted) and $I_{-}$ (deflected) at the first polariser, with the final output beams corresponding to $I_{++}$, $I_{+-}$, $I_{-+}$, and $I_{--}$. Let us consider the fate of beam $I_-$, since the labelling is arbitrary, we must have that $I_{--} = \frac{1}{2}I_0\cos^2(\theta_2-\theta_1) = I_{++}$ to match Malus' law. The only option is then that $I_{+-} = \frac{1}{2}I_0\sin^2(\theta_2-\theta_1) = I_{-+}$. This is also consistent with a Jones matrix analysis of electric field $\vec{E}$. To maintain similarity with the quantum case, our normalised correlation must be $C = \frac{I_{++} + I_{--} - I_{+-} - I_{-+}}{I_0} = \cos(2\theta_2-2\theta_1)$.

\section{Conclusion}\label{sec:conclusion}

In summary, we have shown that there is no coincidence in the fact that Bell correlations match those of a single spin measured successively on different axes. In the latter we measure one spin on two axes, in the former we use entanglement to perform one of those measurements on a member of the pair and perform the second measurement on the other. The inter-pair correlations then allow reconstruction of auto-correlation between axes on the same spin. In a bizarre twist of the fate, given past associations with Bell inequalities~\cite{lambare_bell_2021}, the equivalence of these two experiments proves that quantum mechanics is ``counterfactually definite''. In that, our experiment has the same outcome whether we performed both measurements on one spin or separately on two. Thus, we definitively demonstrate the validity of the counterfactual and dismiss the notion that ``unperformed experiments have no results''~\cite{peres_unperformed_1978}.

The non-locality/contextuality manifested in a Bell scenario now has an obvious source, it is the same as that in the sequential experiment. Thus, they can only be due to the actual physics surrounding local measurements and non-commuting observables. In this case, we argue that the physics at work here is just the effect of quantisation itself. %This is why the GHZ experiment is incompatible with definite values on all spin axes, it is an indirect measurement of quantities that are not actually predefined.

However, this still leaves us with a puzzle. Local hidden variable theories lack some ingredient needed to reproduce quantum mechanics. What we have argued is that this ingredient is neither non-locality nor non-trivial contextuality. The nature of the missing ingredient must be related to the division of observables between those which only reflect pre-determined properties and those in which a state transformation alters the measurement. Indeed, such a division is an essential ingredient in a non-Markovian statistical model that reproduces quantum probabilities~\cite{Barandes-2025}. Perhaps this ingredient is merely quantisation itself, rather than a half-hearted discretisation that is commonly implemented. This question is left for future work.

\backmatter

\bmhead{Acknowledgements}

Thanks to Isaac Nape and Moslem Mahdavifar for their encouragement and helpful conversations.

\bmhead{Declarations}

The author declares funding received from the National Research Foundation of South Africa, under the rated researcher incentive program. The author declares no conflicts of interest.


\bibliography{no_bell.bib}% common bib file

\end{document}